# Optimal $T_c$ for Electron-Doped Cuprate Realized under High Pressure

Shintaro Ishiwata[1]*, Daichi Kotajima[1], Nao Takeshita[2], Chieko Terakura[3], Shinichiro Seki[1], and Yoshinori Tokura[1,3]

[1]*Department of Applied Physics and Quantum-Phase Electronics Center (QPEC), University of Tokyo, Tokyo 113-8656, Japan*
[2]*Electronics and Photonics Research Institute, National Institute of Advanced Industrial Science and Technology, Tsukuba, Ibaraki 305-8568, Japan*
[3]*RIKEN Center for Emergent Matter Science (CEMS), Wako, Saitama 351-0198, Japan*

The race to obtain a higher critical temperature ($T_c$) in the superconducting cuprates has been virtually suspended since it was optimized under high pressure in a hole-doped trilayer cuprate. We report the anomalous increase in $T_c$ under high pressure for the electron-doped infinite-layer cuprate $Sr_{0.9}La_{0.1}CuO_2$ in the vicinity of the antiferromagnetic critical point. By the application of a pressure of 15 GPa, $T_c$ increases to 45 K, which is the highest temperature among the electron-doped cuprates and ensures unconventional superconductivity. We describe the electronic phase diagram of $Sr_{1-x}La_xCuO_2$ to discuss the relation between the antiferromagnetic order and superconductivity.

Electron-hole symmetry/asymmetry has been conceived as one of the most important features of the high-$T_c$ cuprates.[1–3] The asymmetry is manifested in generic phase diagrams as a function of the doping level. The antiferromagnetic (AF) phase exists in a low and narrow doping range for the hole-doped cuprates, whereas that for the electron-doped cuprates remains up to a high doping level, allowing the emergence of superconductivity. However, the electronic phase diagram for the electron-doped cuprates relies only on results for the family of $R_{2-x}Ce_xCuO_4$ (R: rare-earth metal)[4,5], which leaves uncertainty in discussions on the generic electron-hole asymmetry. For an other electron-doped cuprate, infinite-layer $Sr_{1-x}R_xCuO_2$ with the optimum $T_c$ of 43 K, while the superconductivity has been investigated as a function of $x$,[6–9] the relation between the AF order and the superconductivity remains unclear. Since $Sr_{1-x}La_xCuO_2$ contains no magnetic cations other than the Cu ion, the interplay between the magnetism and superconductivity has been extensively studied by such probing methods as nuclear magnetic resonance (NMR), muon spin resonance (µSR), and angle-resolved photoemission spectroscopy (ARPES).[10–12] To extract generic features of electron-doped high-$T_c$ cuprates, a systematic study of $Sr_{1-x}La_xCuO_2$ is highly required. However, this is challenging due to the difficulty in preparing high-quality samples, and thus the pairing symmetry is still under debate.[8,13–15]

*E-mail address: ishiwata@ap.t.u-tokyo.ac.jp

Another important feature with regard to the electron-hole asymmetry can be found in the pressure effect on $T_c$. The hole-doped systems often show an anomalous increase in $T_c$, which has been explained by the change in the carrier concentration,[16] the destabilization of the spin-charge stripe order,[17] and the stabilization of the single-band state through crystal-field splitting.[18] In fact, the highest $T_c$ has been realized under high pressures of above 15 GPa for a Hg-based multilayer cuprate.[19,20] On the other hand, the pressure effect on $T_c$ for the electron-doped cuprates is typically small and negative.[21–24] So far, no improvement of $T_c$ has been observed for the electron-doped infinite-layer cuprates up to high pressures of 8 GPa. Such a remarkable electron-hole asymmetry has been discussed as evidence for conventional BCS-type superconductivity in the electron-doped infinite-layer cuprates.[24]

In this Letter, we have investigated the electronic phase diagram of $Sr_{1-x}La_xCuO_2$ and the pressure effect on $T_c$. The electronic phase diagram suggests a common feature of the electron-doped cuprates, as manifested by the robust AF order competing with the superconductivity. In contrast with the case of $R_{2-x}Ce_xCuO_4$, however, we have observed a significant increase in $T_c$ from 42 to 45 K for $Sr_{0.9}La_{0.1}CuO_2$ under high pressures of up to 15 GPa.

Polycrystalline samples of $Sr_{1-x}La_xCuO_2$ were synthesized by a two-step procedure.[9] First, ambient-pressure-phase $Sr_{1-x}La_xCuO_2$ was synthesized by the following method. Stoichiometric amounts of $SrCO_3$, $CuO$, and $La_2O_3$ were mixed and sintered at 950 °C for 48 h in air. The reacted sample was ground and pressed into pellets, and then sintered again at 950 °C for 24 h. The products were ground into powders and annealed in $N_2$ gas flow for 24 h at 740-770 °C in order to remove excess oxygen. The precursors were placed in a gold capsule with added Ti powder, an oxygen getter, at both ends of the gold capsule and heat treated at 3 GPa and 950-1050 °C for 30 min, followed by quenching to room temperature. A magnetic property measurement system (MPMS, Quantum Design) and physical property measurement system (PPMS, Quantum Design) were used for magnetization and resistivity measurements, respectively. The samples used for transport measurements were cut into a rectangular shape with dimensions of $1.5 \times 1 \times 0.3$ mm$^3$. Resistivity under high pressure was measured by a four-probe method using a cubic-anvil-press apparatus equipped with sintered diamond anvils.[25–27] Daphne 7474 oil was used as a pressure-transmitting medium.

Figure 1(a) shows the temperature dependence of the magnetic susceptibility of Sr$_{1-x}$La$_x$CuO$_2$ measured during heating under a magnetic field of 10 Oe after zero-field cooling. Magnetic shielding effects are observed below 42 and 39 K for the compounds with $x$ = 0.1 and 0.125, respectively, but not for the compound with $x$ = 0.05. The superconducting volume fractions are larger than 50%, indicating bulk superconductivity. Successful electron doping is also confirmed by resistivity measurements as shown in Fig. 1(b). The compound with $x$ = 0.05 shows hopping type conduction without any signature of superconductivity. For $x$ = 0.1 and 0.125, the resistivities are 20 times as low as that for $x$ = 0.05, while still showing the hopping-type behavior in the normal state, presumably reflecting the contribution of incoherent conduction along the $c$ axis. The inset of Fig. 1(b) shows the magnetic field dependence of the Hall resistivity for the compound with $x$ = 0.1, evidencing the predominance of n-type carriers. The carrier concentration of $2.0 \times 10^{21}$ cm$^{-3}$ (corresponding to an electron number of 0.11 per Cu ion) estimated by assuming the single-band model is consistent with the nominal composition ($x$), but not with the band filling ($n$ = 1-$x$), as in all known underdoped high-$T_c$ cuprates.[28] The systematic change in the electronic state by electron doping can also be found in the high-temperature magnetic susceptibility [Figs. 1(c)-1(e)]. It has been reported that the mother compound ($x$ = 0) shows a change in the slope of magnetic susceptibility at 442 K, which was assigned to an AF order in an NMR study.[29] For the compounds with $x$ = 0.05 and 0.1, a similar change in the slope of magnetic susceptibility can be discerned at 320 and 230 K, respectively. Consistent with these results, a μSR study has shown the occurrence of magnetic ordering above room temperature and at 220 K for compounds with $x$ = 0.05 and 0.1, respectively.[11] For $x$ = 0.125, on the other hand, such a kink indicating the onset of AF order is absent from the slope of magnetic susceptibility. It should be noted that the compound with $x$ = 0.1 shows superconductivity with a superconducting volume fraction as large as that of the compound with $x$ = 0.125, while further studies are necessary to determine how the AF order coexists with the superconductivity. Recent ARPES studies for $x$ = 0.1 suggest the intrinsic coexistence of AF ordering and superconductivity,[12] although it remains unclear whether or not the AF order develops in a long range.

The electronic phase diagram of Sr$_{1-x}$La$_x$CuO$_2$ derived from our results and previous reports is shown in Fig. 2(b). As shown in Fig. 2(a), the in-plane lattice constant $a$ changes monotonically with the electron doping level $x$. The AF order is gradually suppressed by electron doping and the quantum critical point appears to be located inside the superconducting dome as reported for $R_{2-x}Ce_xCuO_4$.[4,5] It has been suggested that while doped holes mainly occupy the oxygen $p$ orbitals and cause magnetic frustration effectively suppressing the AF order, doped electrons mainly occupy the Cu $d$ orbitals and quench the local spin moment with less effect on the AF order.[30] Therefore, we conclude that such a robust AF order competing with the superconductivity is a generic feature of the electron-doped high-$T_c$ cuprates.

Next, we show the temperature dependence of the resistivity of $Sr_{0.9}La_{0.1}CuO_2$ under high pressure in Fig. 3(a). While the resistivity ($\rho$) at room temperature decreases monotonically with increasing pressure presumably due to the reduction of resistivity at the grain boundaries, the system does not show a metallic temperature dependence ($d\rho/dT > 0$) even at 15 GPa [see the inset in Fig. 3(a)]. Although the superconducting transitions at high pressures are slightly blurred as compared with that at ambient pressure, a positive shift of $T_c$ with increasing pressure $P$ can be clearly seen between 3 and 15 GPa. $T_c$ at 3 GPa is nearly the same as or slightly lower than that at ambient pressure, consistent with previous reports.[22,23] Above 3 GPa, however, the onset $T_c$ increases from 42 to 45 K at 15 GPa as shown in Fig. 3(b) ($\Delta T_c/\Delta P = 0.25$ K/GPa); $T_c = 45$ K is the highest temperature among the electron-doped cuprates and is expected to become even higher with the further application of pressure.

Let us discuss the possible origin of the pressure-induced enhancement of $T_c$ in $Sr_{0.9}La_{0.1}CuO_2$. Since $T_c$ is almost constant as a function of $x$, the change in the carrier concentration can be ruled out as the origin of the positive pressure effect. It has been reported for $Nd_{2-x}Ce_xCuO_4$ that while hydrostatic pressure has no effect on $T_c$,[21] uniaxial pressure along the $c$ axis causes a significant decrease in $T_c$. Therefore, in-plane compression, which strengthens the AF exchange coupling, is expected to enhance $T_c$ for $Nd_{2-x}Ce_xCuO_4$.[31] Likewise for $Sr_{0.9}La_{0.1}CuO_2$, the positive pressure effect on $T_c$ can be ascribed to the predominance of the in-plane compression effect enhancing the AF exchange coupling. The difference in the pressure effect on $T_c$ between these two families of electron-doped cuprates may arise from the structural difference in the block-layer units, i.e. the $Sr_{1-x}La_x\square$ layer ($\square$ represents oxygen vacancy) vs. the $Nd_{2-x}Ce_xO_2$ layer.

In summary, the highest onset $T_c$ in the electron-doped cuprates was revealed in infinite-layer $Sr_{0.9}La_{0.1}CuO_2$ under high pressure. The application of pressure of up to 15 GPa enables us to clearly observe the positive pressure effect on $T_c$. Since the compound with $x =$

0.1 resides on the verge of the quantum critical point, the AF correlation is expected to play an important role in the superconductivity, thus leading to the positive pressure effect on $T_c$ through in-plane compression. The proximity of the superconducting dome to the quantum critical point and the pressure-induced increase in $T_c$ to above 40 K are indicative of the $d$-wave pairing mechanism in this system, as recently suggested by phase-sensitive measurements.[15] This work highlights the potential of the electron-doped cuprates as high-$T_c$ superconductors to be further optimized.

**Acknowledgments**

The authors thank R. Arita and T. Sasagawa for valuable discussions and M. Azuma for his advice on sample preparation. This study was partly supported by a Grant-in-Aid for Young Scientists (Grant No. 23685014) from MEXT and by the Funding Program for World-Leading Innovative R&D on Science and Technology (FIRST Program), Japan.

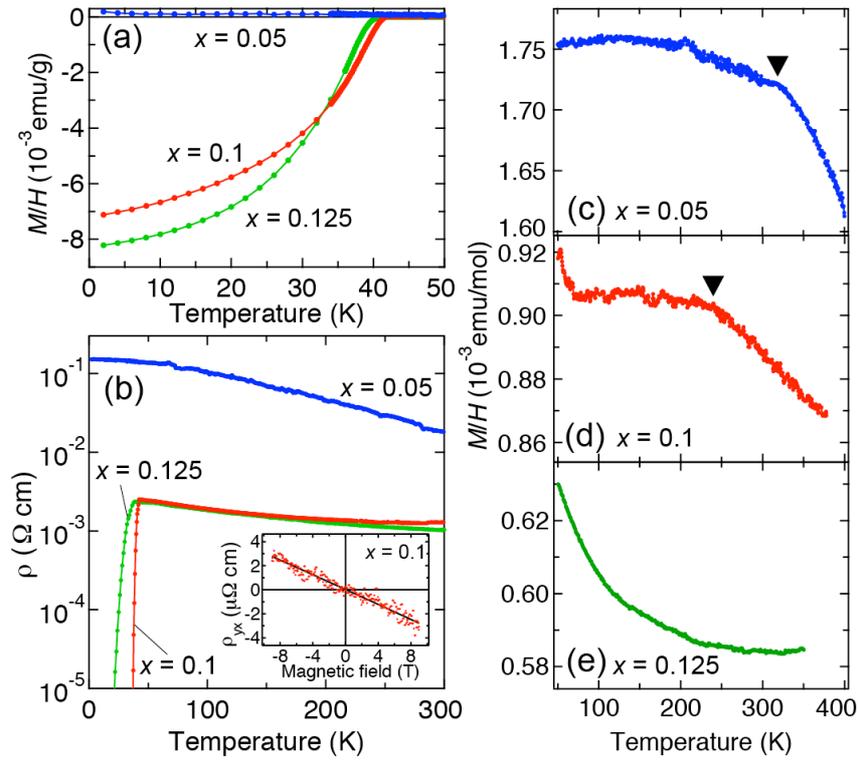

Fig. 1. (Color online) Electronic and magnetic properties of $Sr_{1-x}La_xCuO_2$. Temperature dependence of (a) magnetic susceptibility under an applied magnetic field of 10 Oe, (b) resistivity, and (c)-(e) magnetic susceptibility measured with a magnetic field of 1000 Oe at high temperatures. The inset in (b) shows Hall resistivity as a function of magnetic field for $x = 0.1$ at 100 K. The triangles in (c) and (d) indicate the Néel temperature.

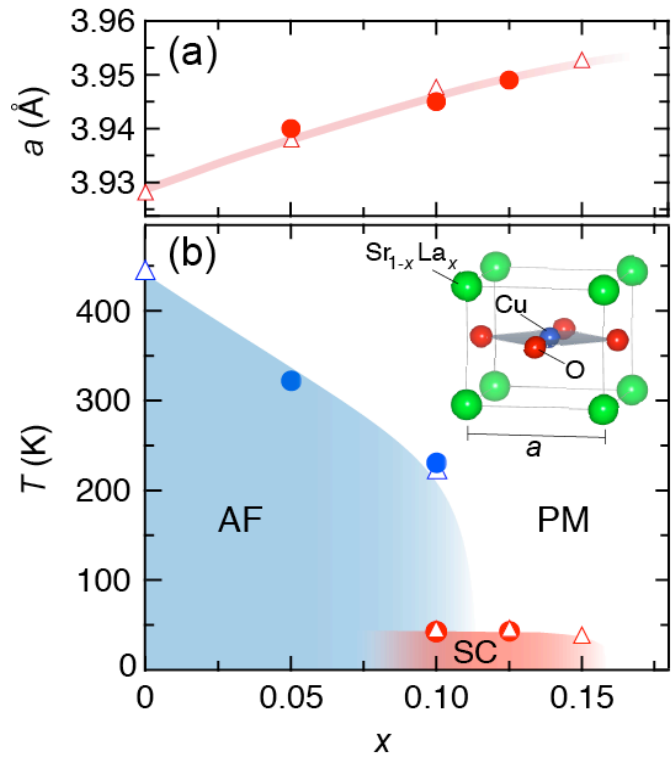

Fig. 2. (Color online) Variation of lattice and electronic state of $Sr_{1-x}La_xCuO_2$ with electron doping. (a) In-plane lattice constant $a$. The triangles were reproduced from ref. 6. (b) Electronic phase diagram. PM, AF, and SC represent paramagnetic, antiferromagnetic, and superconducting phases, respectively. The triangles for the Néel temperature and superconducting transition temperature were taken from previous reports. [8, 11, 29] The inset shows the crystal structure.

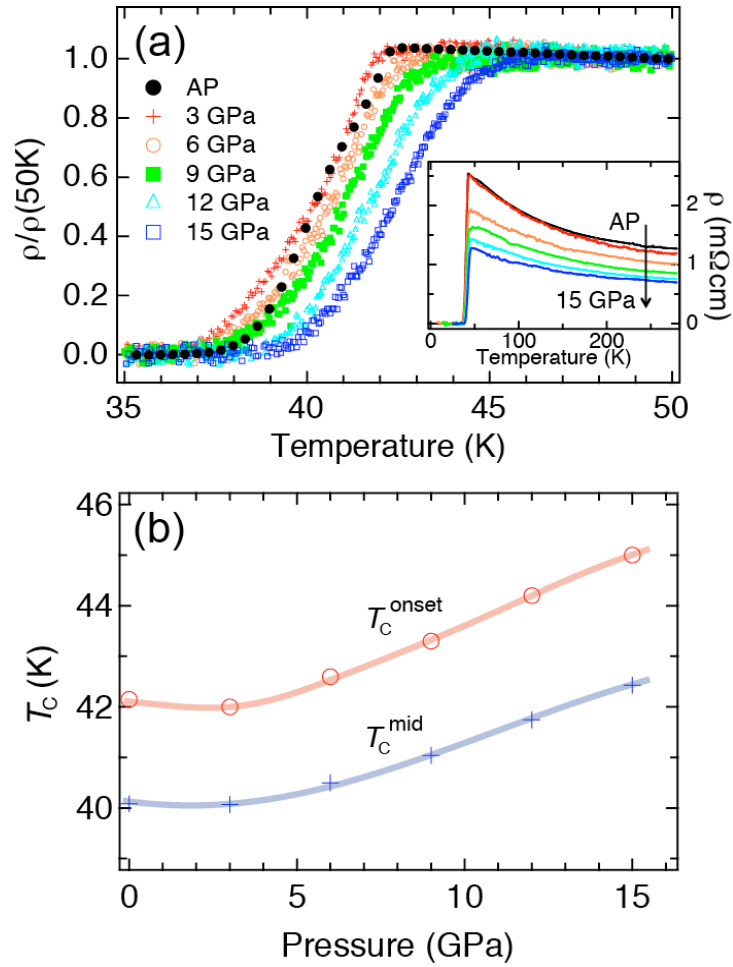

Fig. 3. (Color online) Pressure effect on $T_c$ for $Sr_{0.9}La_{0.1}CuO_2$. (a) Temperature dependence of normalized resistivity defined by $\rho(T) = \rho(50\ K)$ under high pressures of up to 15 GPa. Resistivity at ambient pressure (AP) was measured without the pressure cell. The inset shows resistivity as a function of temperature under high pressure. (b) Pressure dependence of $T_c$ of $Sr_{0.9}La_{0.1}CuO_2$. The circles show the onset $T_c$. The crosses show the temperature $T_c^{mid}$ at which the slope of the resistivity is maximized.